\documentclass[aps,prl,amsfonts,amssymb,epsfig,twocolumn,superscriptaddress,showpacs]{revtex4}
\usepackage{color}
\usepackage{graphicx}
\usepackage{epstopdf}
\definecolor{darkblue}{rgb}{0,0,.8}
\usepackage{bm}

\begin{document}
\title{Trap modulation spectroscopy of the Mott-insulator transition in optical lattices}

\author{H. Lignier}
\affiliation{CNR-INFM, Dipartimento di Fisica `E. Fermi', Largo Pontecorvo 3, 56127 Pisa, Italy}
\author{A. Zenesini}
\affiliation{CNR-INFM, Dipartimento di Fisica `E. Fermi', Largo
Pontecorvo 3, 56127 Pisa, Italy} \affiliation{CNISM UdR
Universit\`a di Pisa, Largo Pontecorvo 3, 56127 Pisa, Italy}
\author{D. Ciampini}
\affiliation{CNR-INFM, Dipartimento di Fisica `E. Fermi', Largo
Pontecorvo 3, 56127 Pisa, Italy} \affiliation{CNISM UdR
Universit\`a di Pisa, Largo Pontecorvo 3, 56127 Pisa, Italy}
\author{O. Morsch}
\affiliation{CNR-INFM, Dipartimento di Fisica `E. Fermi', Largo Pontecorvo 3, 56127 Pisa, Italy}
\author{E. Arimondo}
\affiliation{CNR-INFM, Dipartimento di Fisica `E. Fermi', Largo Pontecorvo 3, 56127 Pisa, Italy}
\author{S. Montangero}
\affiliation{Institut f\"ur Quanteninformationsverarbeitung, Universit\"at Ulm, D-89069 Ulm, Germany }
\affiliation {NEST CNR-INFM \& Scuola Normale Superiore, Piazza dei Cavalieri 7, I-56126 Pisa, Italy}
\author{G. Pupillo}
\affiliation{Institute for Theoretical Physics, University of Innsbruck,  A-6020 Innsbruck, Austria}
\affiliation{Institute for Quantum Optics and Quantum Information, A-6020 Innsbruck, Austria}
\author{R. Fazio}
\affiliation {NEST CNR-INFM \& Scuola Normale Superiore, Piazza dei Cavalieri 7, I-56126 Pisa, Italy}

\begin{abstract}
We introduce a new technique to probe the properties of an
interacting cold atomic gas that can be viewed as a dynamical
compressibility measurement. We apply this technique to the study
of the superfluid to Mott insulator quantum phase transition  in
one and three dimensions for a bosonic gas trapped in an optical
lattice. Excitations of the system are detected by time-of-flight
measurements. The experimental data for the one-dimensional case
are in good agreement with the results of a time-dependent density
matrix renormalization group calculation.
\end{abstract}


\maketitle It is now possible to study strongly interacting
condensed matter systems using cold atoms trapped in optical
lattices~\cite{lewenstein,bloch_review}. Recent experiments have
led to the  realization of  a Tonks-Girardeau gas in one-dimension
(1D)~\cite{paredes04,kinoshita04} and to the detection of the
quantum phase transition from a superfluid (SF) to a
Mott-insulator (MI) in bosonic
systems~\cite{greiner02a,stoferle_04,spielman_07}, to name just
two. Equilibrium (ground state) properties of the various
condensed phases can be extracted from the interference pattern of
the gas after expansion (see for example~\cite{greiner02a}).
Dynamical properties, however, cannot be characterized in this
way. In order to have access to the excitation spectrum, and
therefore to the dynamical properties, additional spectroscopic
tools such as Bragg spectroscopy~\cite{stenger99} have been
designed. St\"{o}ferle {\em et al.}~\cite{stoferle_04} introduced
modulation spectroscopy based on a periodic change of the optical
lattice depth and analyzed the energy absorption using
time-of-flight techniques.

In this Letter we introduce a new technique which enables a direct
probing of the phase diagram and excitation spectrum of an
interacting cold atomic gas. Excitations in the system are
produced by periodically modulating the strength of an external
harmonic potential that confines the atoms. The width of the
expanded atomic cloud in time-of-flight experiments is then used
as a measure of the increase in energy of the system due to the
trap modulation. We use this technique to study the superfluid to
Mott insulator quantum phase transition obtained upon increasing
the depth of the optical lattice. We observe a noticeably
different behavior in the 1D and 3D results. While the 3D results
show a sharp transition consistent with mean-field predictions,
the 1D results show a broad transition between the superfluid and
insulating phases. A comparison with exact time-dependent DMRG
results in 1D for a model system allows us to identify the region
of formation of a "wedding-cake" structure in the atomic density
profile in 1D, with coexisting superfluid and insulating regions.
The latter has been recently observed experimentally in
3D~\cite{shell_ketterle_bloch_06}.

The trap modulation spectroscopy presented here can be interpreted
as a {\em dynamical} compressibility measurement, performed in a
regime where the modulation frequency is much larger than the
inter-well tunneling frequency. This is in contrast to recent
proposals of measuring the compressibility of the system in the
sense of a volume reduction following an adiabatic
compression~\cite{roscilde08}, as recently performed in a system
of interacting fermions~\cite{Schneider}. We show below that our
technique is well suited to characterizing dynamical quantities
such as, e.g., the change of the excitation spectrum across the
superfluid to Mott insulator quantum phase transition.

\begin{figure}[ht]
\includegraphics[width=8cm]{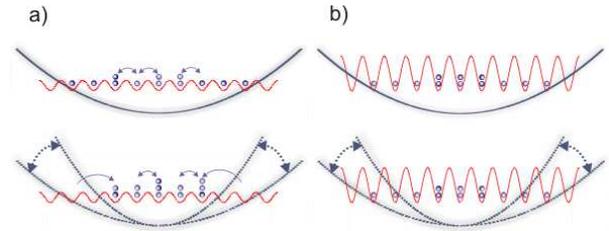}
\caption{\label{figure1}(color online) Trap modulation
spectroscopy of a Bose gas in an optical lattice. a) In the
superfluid regime repeated compressions of the trap lead to a
compression of the Bose gas and hence to excitations, whereas in
the Mott insulator regime (b) the Bose gas becomes incompressible
and no excitation takes place. In a) and b) the upper diagram
represents the initial condition before applying the modulation.}
\end{figure}

The working principle of our method is illustrated in Fig.~1. A
Bose-Einstein gas is held in the combined potential of a harmonic
trap of mean frequency $\omega$ and an optical lattice of depth
$V_0$. The frequency of the harmonic trap is then increased and
decreased in a sinusoidal fashion, with the instantaneous mean
trap frequency given by
\begin{equation}
\omega(t) = \omega+ \delta\omega\sin^2(\Omega t/2),
\end{equation}
where $\delta \omega$ is the peak-to-peak modulation depth and
$\Omega$ the modulation frequency. Figure 2 shows results of exact
time-dependent density-matrix-renormalization-group (tDMRG)
calculations of this scheme for a model system of $N=15$ particles
in one dimension. The system's dynamics is modeled using the
Bose-Hubbard (BH) Hamiltonian~\cite{jaksch}, with a second order
Trotter expansion of the Hamiltonian and time-steps of
$0.01J$~\cite{daley04} (where J is the inter-well tunneling
energy), in which we take advantage of the conserved total number
of particles $N$ by projecting onto the corresponding subspace.
The truncated Hilbert space dimension is up to $m = 100$, while
the allowed number of particles per site is $D = 8$. The trap
modulation Eq.(1) is accounted for by a term $\frac{1}{2}
m\omega(t)^2 d_L^2\sum_i i^2 \hat{n}_i$  in the BH-Hamiltonian,
where $m$ is the atomic mass,$d_L$ the lattice constant and
$\hat{n}_i$ the number operator on site $i$~\cite{montangero}. The
BH-Hamiltonian describes well the microscopic dynamics in the
experiment (see below), and Fig. 2 shows the salient features of
the physical picture described above. In the superfluid regime
(Fig. 2(a)), after $8$ modulation cycles of the trap frequency the
distribution of atoms inside the lattice is markedly different
compared to the initial situation as the trap modulation has led
to a compression of the superfluid towards the central lattice
sites. In the Mott insulator regime, on the other hand, the
initial and final distributions are indistinguishable from each
other, demonstrating that the system has not been excited by the
repeated trap compressions.

Our modulation scheme allows us to detect signatures of the
superfluid to Mott insulator transition. Figure~3 (a) shows the
results of the 1D numerical simulation for the relative energy
growth rate during the trap modulation (obtained from a linear fit
to the energy increase for small $t$ and rescaled by the energy at
$t=0$). The relative growth rate data are plotted as a function of
the Hubbard parameter $U/(2dJ)$ for different occupation numbers
$n$ at the central lattice site (typically $n=2$ in our
experiments). A weighted average over different occupation numbers
was also performed in order to take into account the contributions
of different 1D tubes in the deep 2D lattice. The figure shows
that the relative growth rate for $N=15$ and $N=10$ (triangles and
open squares, respectively) changes rapidly in the parameter
region $2\leq U/(2dJ)\leq 8$. We find that this corresponds to the
formation of a "wedding-cake" structure in the density profile
upon increasing $U/(2dJ)$, as in Fig. 2(a) and (b). This
restructuring of the density profile is accompanied by sharp
features in the relative growth rate for $N=15$ and $N=10$, which
are largely washed out by the numerical averaging over the tubes
(black dots).

\begin{figure}[ht]
\includegraphics[width=7cm]{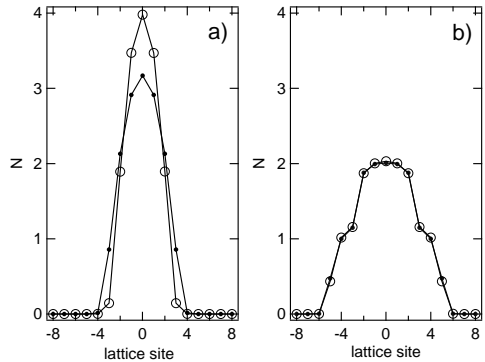}
\caption{\label{figure2} Numerical simulation of the effects of
trap modulation in a one-dimensional lattice with total atom
number $N=15$ (corresponding to $n=2$) for (a) $U/(2dJ)=2$ and (b)
$U/(2dJ)=9$. The filled circles represent the atomic distribution
at $t=0$, while the open circles show the distribution after $8$
compression cycles with $\delta\omega/\omega=1$.}
\end{figure}

In order to test these predictions experimentally, we created
Bose-Einstein condensates of roughly $6\times 10^4$ rubidium-87
atoms inside a crossed dipole trap~\cite{sias_07} of mean
frequency $\omega\approx 2\pi\times 80\,\mathrm{Hz}$. A
three-dimensional optical lattice with lattice constant
$d_L=421\,\mathrm{nm}$ was then adiabatically (within
$100\,\mathrm{ms}$) superimposed on the condensate. The final
depth $V_0$ of the lattice (in the range between
$6\,E_\mathrm{rec}$ and $22\,E_\mathrm{rec}$, where the recoil
energy $E_\mathrm{rec}=\frac{\hbar^2\pi^2}{2md_L^2}$) determined
the Hubbard parameter $U/(2dJ)$, where $U$ is the on-site
interaction energy~\cite{footnote_theory}. We realized the
superfluid to Mott insulator transition both in 1D ($d=1$) and 3D
($d=3$). For the 1D case we ramped two of lattice beams to a
maximum depth of $26\,E_\mathrm{rec}$ (which resulted in a 2D
array of one-dimensional tubes~\cite{greiner_01,stoferle_04}) and
varied the depth $V_0$ of the third lattice, whereas for the 3D
case all three lattices had the same depth. In both cases we
performed various tests (visibility of the interference
pattern~\cite{gerbier_05b}, excitation
spectrum~\cite{stoferle_04}, adiabaticity of the lattice
ramps~\cite{gericke_07}) in order to identify the critical lattice
depth for entering the Mott insulating regime and to ensure that
the various regimes were reached adiabatically, i.e., without
exciting the Bose gas.

Once the final lattice depth was reached, the harmonic trap
frequency was modulated by sinusoidally modulating the power of
the crossed dipole trap between the extreme values $P_0$ and
$P_0(1+\alpha)$ at frequency $\Omega$~\cite{footnote1D}. The
resulting variation of the trap frequency $\omega(t)\propto
\sqrt{P(t)}$ was reasonably sinusoidal for $\alpha\lesssim 1$,
with a modulation parameter $\delta \omega /\omega =
\sqrt{1+\alpha}-1$. After a few milliseconds the modulation was
switched off and the optical lattices were ramped down in
$15\,\mathrm{ms}$ to $V_0=4\,E_\mathrm{rec}$. The atoms were
allowed to thermalize for a further $10\,\mathrm{ms}$, then both
the lattice and the dipole trap were suddenly switched off, and
the atoms were imaged after a time-of-flight of
$23.3\,\mathrm{ms}$. For a fixed lattice depth in the superfluid
regime we observed a roughly linear increase with modulation time
in the width of the expanded Bose gas for up to $10\,\mathrm{ms}$
of trap modulation, and this was independent of the modulation
frequency $\Omega$ over a wide range (between $\approx
0.5\,\mathrm{kHz}$ and $\approx 4\,\mathrm{kHz}$). These findings
are consistent with our numerical simulations.

\begin{figure}[ht]
\includegraphics[width=9cm]{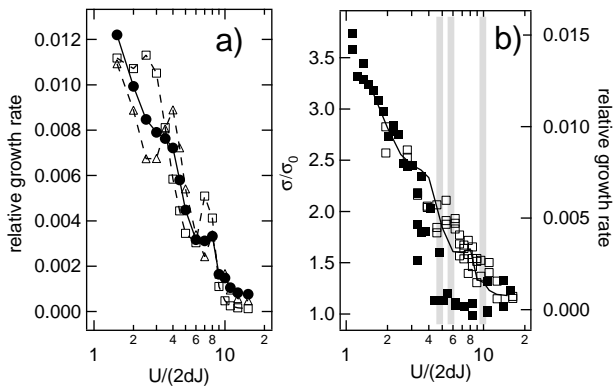}
\caption{\label{figure3}(a) Numerical simulation of trap
modulation spectroscopy in 1D with $\delta \omega / \omega=0.22$
and $\Omega/2\pi=1\,\mathrm{kHz}$. Shown here are the results for
boson numbers $N=15$ (open triangles) and $N=10$ (open squares),
corresponding to occupation numbers $n=2$ and $n=1.5$ of the
central lattice site, respectively, as well as a weighted average
(filled circles) over total atom numbers $N=5$, $N=10$ and $N=15$.
The energy growth rates are extracted by taking the average of two
extreme linear fits to the initial growth curve, leading to a
relative error of around $15$ percent. (b) Experimental results in
1D (open squares) for $\delta \omega / \omega=0.22$ and a
modulation time $t_\mathrm{mod}=5\,\mathrm{ms}$ and 3D (filled
squares) with $\delta \omega / \omega=0.41$ and
$t_\mathrm{mod}=10\,\mathrm{ms}$. For comparison, the averaged
numerical results of Fig. 3 (a) are also shown (solid line). The
grey bars indicate (from left to right) the critical values of
$U/(2dJ)$ for the formation of the first Mott insulator shell
according to Quantum Monte Carlo
calculations~\cite{capogrossosansone07} and mean-field theory, and
for the second shell according to mean-field theory,
respectively.}
\end{figure}

The experimental results for the 1D and 3D cases are shown in Fig.
3 (b), where we plot the width $\sigma$ (normalized to the width
$\sigma_0$ in the absence of any modulation) of the Bose gas in
time-of-flight (which reflects the increase in energy of the
system) after a fixed modulation time $t_\mathrm{mod}$ as a
function of the Hubbard parameter $U/(2dJ)$ for $d=1$ and $d=3$.
In the one-dimensional case, the experimental results are
consistent with the (averaged) 1D numerical simulations shown in
Fig. 3 (a) if an overall scaling factor is applied to the
theoretical results. Generally, we see a decrease in the
excitability of the Bose gas reflected by $\sigma/\sigma_0$
approaching unity. While in 1D this decrease is gradual and even
for large values of $U/(2dJ)$ there is a residual excitation of
the Bose gas, in 3D we observe a sharp drop in $\sigma/\sigma_0$
around $U/(2dJ)=4$. For $U/(2dJ)\gtrsim 5$, $\sigma/\sigma_0$
remains constant around $1$, indicating that the system is
incompressible. This agrees well with the physical picture of a
transition from a compressible (and hence excitable) superfluid to
an incompressible Mott insulator for $U/(2dJ)\approx 5.8$
(mean-field prediction) or $U/(2dJ)\approx 4.9$ (Quantum
Monte-Carlo simulation)~\cite{bloch_review}.

Performing the above experiment in 3D for different values of the
modulation depth $\delta\omega/\omega$, we found that the sharp
feature around $U/(2dJ)=4$ was clearly visible for
$\delta\omega/\omega\lesssim 0.5$ but became increasingly washed
out for larger values. Indeed, for an intermediate value
$\delta\omega/\omega=0.7$ (see Fig.~4 (a)) we saw a structure with
two apparent "steps" at $U/(2dJ)\approx 4$ and $U/(2dJ)\approx 9$,
and $\sigma/\sigma_0$ approached unity for $U/(2dJ) > 10$. The
proximity of the step positions to the theoretical values of
$U/(2dJ)$ for the formation of Mott insulator shells with $n=1$
and $n=2$ atoms per lattice site, respectively,
 \cite{capogrosso,shell_ketterle_bloch_06} suggests the following intuitive
explanation. For small modulation depths, the formation of the
$n=1$-shell greatly reduces the overall excitability of the
system, while for larger values of $\delta\omega/\omega$ the $n=1$
shell, being located away from the trap center and hence exposed
to larger potential gradients when the trap is compressed, remains
partly compressible and only when the $n=2$ shell is formed does
the system become insensitive to the trap modulation.

We tested this hypothesis by measuring the Bose gas response as
a function of $\delta\omega/\omega$ in the superfluid and in the
Mott insulator regime. Figure 4 (b) shows that while in the
superfluid regime $\sigma/\sigma_0$ increased roughly linearly
with the modulation depth, in the Mott insulator regime we
observed that for $\delta\omega/\omega\lesssim 0.6$ there was no
dependence on the modulation depth. For larger values of
$\delta\omega/\omega$, the Bose gas could again be excited,
suggesting that in this regime the trap compression was sufficient
to overcome the potential barrier $U$ for moving atoms between
adjacent lattice sites. Alternatively, a strong trap modulation
might lead to a heating of the thermal fraction of the sample,
thus "melting" the Mott insulator \cite{gerbier07}  and increasing
the compressibility of the Bose gas . While we were unable to
rule out either of these two scenarios, this remains an
interesting question for future experiments and theoretical
analysis.

\begin{figure}[ht]
\includegraphics[width=9cm]{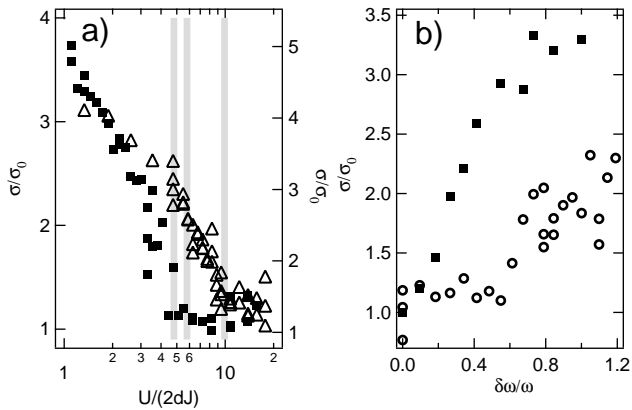}
\caption{\label{figure4} (a) Trap modulation spectroscopy in 3D
for $\delta\omega/\omega=0.22$ (filled squares) and
$\delta\omega/\omega=0.7$ (open triangles). The grey bars are
defined in Fig.~3. (b) The response of the Bose gas to trap
modulation as a function of the modulation depth
$\delta\omega/\omega$ for $U/(2dJ)=2$ (filled squares) and
$U/(2dJ)=9$ (open circles).}
\end{figure}

Finally, we investigated the response of our system to trap
modulation for adiabatic and non-adiabatic loading of the lattice.
In all of the experiments described above, we made sure that the
timescales and shapes of the intensity ramps for the lattice beams
were such that we reached the final lattice depth adiabatically
and hence always prepared the system in its ground state. One can,
however, prepare the same final lattice depth non-adiabatically,
in which case one expects to find the system in an excited state.
If the lattice depth is above the critical value for the Mott
insulator transition, this means that for non-adiabatic loading
the system will {\it not} be in a pure Mott insulator state and
should, therefore, retain its compressibility. We tested this
assumption by using a two-part ramp for the optical lattice with a
$100\,\mathrm{ms}$ exponential ramp up to $5\,E_\mathrm{rec}$ and
a linear ramp in $0.1\,\mathrm{ms}$ to the final value $V_0$. The
second part of the ramp was highly non-adiabatic, which we
verified by lowering the lattice after a holding time of
$5\,\mathrm{ms}$ and checking that the Bose gas was excited, as
expected. We then performed trap modulation spectroscopy using a
short modulation time of $5\,\mathrm{ms}$ in order to minimize the
role of the spurious effects on $\sigma/\sigma_0$ due to the
loading process. Figure~5 shows that for a non-adiabatic loading
ramp, the Bose gas does, indeed, remain excitable well above the
critical value for the formation of a Mott insulator.

\begin{figure}[ht]
\includegraphics[width=7cm]{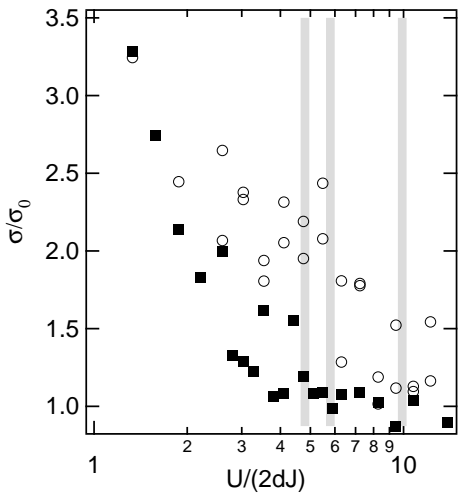}
\caption{\label{figure5} Trap modulation spectroscopy for an
adiabatically loaded lattice (filled squares) and for
non-adiabatic loading (open circles). In this experiment,
$\delta\omega/\omega=0.41$, and the grey bars are defined in
Fig.~3.}
\end{figure}

In summary, we used trap modulation spectroscopy in order to study
the superfluid to Mott-insulator phase transition. The method
presented here seems to be a promising tool for the
characterization of strongly interacting cold atomic systems. For
the one-dimensional case we obtained good agreement with exact
numerical calculations. For the three-dimensional case a full
theoretical interpretation will require further calculations. As a
perspective for future studies we believe that it is of great
importance to explore in detail the differences and similarities
of our approach with other methods based on a quasi-static
compression of the trap~\cite{Schneider}.

Financial support by MIUR PRIN-2007 and the E.U. projects SCALA
and NAMEQUAM is greatfully acknowledged. Numerical simulations in
this work were done using the DMRG code released within the PwP
project (www.dmrg.it).

\bibliographystyle{apsrmp}

\begin{thebibliography}{99}
\bibitem{footnote_theory} We note here that while in our experiment
$J$ is changed through $V_0$ (which has little effect on $U$ in
the range of parameters used here), in the numerical simulation
$J$ is kept constant and $U$ is varied. The latter situation could
be exactly reproduced experimentally by using a Feshbach
resonance. The agreement between our experimental results and the
simulation indicates that it is the ratio $U/J$ rather than the
two individual parameters that governs the system's behaviour.
\bibitem{footnote1D} In the 1D experiments we checked that the
trap modulation in the direction perpendicular to the tubes (i.e.
in the direction of the 2D lattice) did not lead to an excitation
of the Bose gas for any value of the lattice depth of the 1D
lattice. This was done by modulating only the dipole trap beam
that was approximately collinear with the 1D lattice beam and
checking the response of the Bose gas as a function of $V_0$.
\bibitem{capogrossosansone07} Deviations from the mean-field thresholds are discussed in \cite{bloch_review}.
\end{thebibliography}

\end{document}